# Surface Albedo and Spectral Variability of Ceres

Jian-Yang Li (李荐扬)[1], Vishnu Reddy[1], Andreas Nathues[2], Lucille Le Corre[1], Matthew R. M. Izawa[3,4], Edward A. Cloutis[3], Mark V. Sykes[1], Uri Carsenty[5], Julie C. Castillo-Rogez[6], Martin Hoffmann[2], Ralf Jaumann[5], Katrin Krohn[5], Stefano Mottola[5], Thomas H. Prettyman[1], Michael Schaefer[2], Paul Schenk[7], Stefan E. Schröder[5], David A. Williams[8], David E. Smith[9], Maria T. Zuber[10], Alexander S. Konopliv[6], Ryan S. Park[6], Carol A. Raymond[6], Christopher T. Russell[11]

…

[1] Planetary Science Institute, 1700 E. Ft. Lowell Rd., Suite 106, Tucson, AZ 85719, USA

[2] Max Planck Institute for Solar System Research, 37077 Göttingen, Germany

[3] University of Winnipeg, Winnipeg, Canada

[4] Royal Ontario Museum, Toronto, Canada

[5] German Aerospace Center (DLR), Institute of Planetary Research, Berlin, Germany

[6] Jet Propulsion Laboratory, California Institute of Technology, Pasadena, CA 91109, USA

[7] Lunar and Planetary Institute, Houston, TX 77058, USA

[8] School of Earth and Space Exploration, Arizona State University, Tempe, AZ 85287, USA

[9] Solar System Exploration Division, NASA Goddard Space Flight Center, Greenbelt, MD 20771, USA

[10] Department of Earth, Atmospheric and Planetary Sciences, Massachusetts Institute of Technology, Cambridge, MA 02139, USA

[11] Institute of Geophysics and Planetary Physics, University of California, Los Angeles, CA 90095, USA

**Abstract**

Previous observations suggested that Ceres has active but possibly sporadic water outgassing, and possibly varying spectral characteristics in a time scale of months. We used all available data of Ceres collected in the past three decades from the ground and the Hubble Space Telescope, and the newly acquired images by Dawn Framing Camera to search for spectral and albedo variability on Ceres, in both a global scale and local regions, particularly the bright spots inside Occator crater, over time scales of a few months to decades. Our analysis has placed an upper limit on the possible temporal albedo variation on Ceres. Sporadic water vapor venting, or any possibly ongoing activity on Ceres, is not significant enough to change the albedo or the area of the bright features in Occator crater by >15%, or the global albedo by >3% over various time scales that we searched. Recently reported spectral slope variations can be explained by changing Sun-Ceres-Earth geometry. The active area on Ceres is less than 1 km$^2$, too small to cause global albedo and spectral variations detectable in our data. Impact ejecta due to impacting projectiles of tens of meters in size like those known to cause observable changes to the surface albedo on Asteroid Scheila cannot cause detectable albedo change on Ceres due to its relatively large size and strong gravity. The water vapor activity on Ceres is independent of Ceres' heliocentric distance, ruling out the possibility of comet-like sublimation process as a possible mechanism driving the activity.

1. Introduction

Observations of dwarf planet Ceres by NASA's Dawn spacecraft have revealed a surface peppered with bright spots associated with impact craters (Nathues et al. 2015). The bright spots are prominent albedo features on the otherwise uniform surface of Ceres in albedo and color. Compositional analysis reveals that these bright spots are likely made up of a mixture of water ice, salts, and dark background material, suggesting a briny subsurface source. Bright spots within impact craters Dantu (125 km) and Occator (92 km) have been linked to intermittent and localized water vapor sources observed by the Herschel Space Observatory in 2011-2013 (Küppers et al. 2014), and therefore possibly related to the sublimation activity. Recent ground-based observations of Ceres also suggested short-term global spectral variability that was attributed to changing amount of water ice on the surface (Perna et al. 2015). These observations suggest a possibly changing surface caused by water outgassing on time scales of decades to months. We searched for evidence of surface albedo and spectral variations that might be evidence for ongoing activity by comparing various historical datasets and the most recent Dawn Framing Camera (FC) data collected during the approach to Ceres. Our search covered three different time frames: long-term in decades; mid-term in years, and short-term in months.

2. Analysis

*2.1 Full-disk spectral variability*

The spectral data archived in the Planetary Data System Small Bodies Node from four epochs (Vilas et al. 1998, Bus & Binzel 2003, Lazzaro et al. 2006), from Perna et al. (2015) in two epochs, and from recent ground-based observations (Reddy et al. 2015a) span over nearly three decades, suitable for searching for long-term spectral variability of Ceres (Fig. 1a). The spectra of Ceres from various datasets all show a nearly linear shape between 0.53 and 0.85 µm, with increasing spectral slopes with solar phase angle, and absorptions <0.54 µm and >0.9 µm. Therefore we fit linear slopes to all spectra between 0.54 and 0.85 µm to quantify the spectral shape variation. Fig. 1b suggests that the variations in slope are ascribable to variations in phase angle, except for the Dawn RC2 observations at 44º phase angle (described below), possibly indicating that the approximate linear relationship between spectral slope and phase angle stops somewhere between 20º and 40º phase angle.

Dawn FC observed Ceres for at least one full rotation of Ceres of 9.075 hours Chamberlain et al. 2007) in three separate epochs in February and April 2015 during the approach. These three observations are named Rotational Characterization (RC1, RC2, and RC3, see Table 1). The disk of Ceres did not fill the field of view of the FC during the RCs. Image calibration and spectral cube computation have been performed according to Reddy et al. (2012a) and Nathues et al. (2014). The time difference between RC1 and RC2 is one week and between RC2 and RC3 ten weeks, enabling us to explore short-term (days/weeks/months) global spectral variability on Ceres. To minimize dependence on the effect of the absolute calibration of the instrument, we calculated the relative spectra in each of the three RCs. The phase angle of RC3 is about 8º. Based on the linear fit to the spectral slope with respect to phase angle (Fig. 1b), the spectral slope of Ceres at the RC3 phase angle is near zero. Therefore, the ratio spectra of RC1 and RC2 to RC3 should remove the effect of the uncertain absolute radiometric calibration of Dawn FC (Schröder et al. 2013a, 2014), and approximate the spectra of Ceres collected at RC1 (17º phase angle) and RC2 (44º). The spectral slope from RC1 is consistent with the overall trend with phase angle in the three RCs. We therefore based our spectral slope analysis of Dawn FC data on these relative spectra (Fig. 1a).

*2.2 Global photometric variability*

To measure global variations in the albedo of Ceres over a time scale of years, we used previously published HST data collected in 2003 and 2004 with the Advanced Camera for Survey (ACS) High Resolution Channel (HRC) through the broadband F555W filter (Li et al. 2006), supplemented by observations taken with the Wide Field Camera 3 (WFC3) UVIS channel through the broadband filter F555W in 2014 (GO-13503, PI: B.E. Schmidt). The WFC3/UVIS F555W has a similar combined throughput as the ACS/HRC F555W filter in 2014. We scaled the total flux of Ceres to a phase angle of 6º by a linear fit to the disk-integrated phase function measured from the HST ACS/HRC data, which span a small range of phase angles from 5.4º to 7.7º, and phasing each flux measurement to the rotation of Ceres based on the rotational period of Ceres, which has an accuracy of 7 ms per rotation (Chamberlain et al. 2007). The total reflected flux of Ceres measured through the three datasets agrees with each other within 3%, well within the absolute photometric calibration uncertainties of the two instruments.

*2.3 Occator bright spots*

The HST/ACS/HRC data and Dawn FC data are separated by 12 years, and are both spatially resolved, suitable for searching for albedo changes in local scales on Ceres in a medium timeframe of years. We focus on the Occator crater because the bright spots inside this crater are undoubtedly the most prominent features on Ceres, and possibly related to the water vapor activity (Nathues et al. 2015). Comparison of absolute feature brightness between different datasets is extremely difficult, due primarily to the vastly different optical characteristics and absolute calibrations between HST and FC. Our approach is to compare the Occator bright spots with a reference region, for which we chose the Haulani crater (32 km diameter), because it is relatively easy to identify in HST images. The underlying assumptions for our comparative approach is that the reflectances of Occator bright spots and Haulani crater did not change simultaneously, and this is justified by the unchanging disk-integrated brightness of Ceres over the same time period (§2.2). We used RC1 images and RC3-equator images (Table 1) in this study. Note that Haulani crater is in fact the brightest feature on Ceres in HST images (Li et al. 2006, Thomas et al. 2005), brighter than Occator crater (Fig. 2). The present study seeks an answer to the question: Did the brightness of Occator crater change since the previous HST observations, or are the different relative reflectance values of Haulani and Occator an observing artifact from comparing the HST and Dawn datasets?

*2.3.1 Photometric modeling*

Precise photometric comparisons require corrections for different illumination and viewing geometries, which further rely on photometric models of the whole surface of Ceres and the bright spots. We separately modeled the photometric properties of Occator bright spots and the global average of Ceres based on RC3 clear filter (effective wavelength 730 nm, FWHM 370 nm) data at 1.3 km/pixel with the Hapke model (Hapke 2012), following the procedure outlined in Li et al. (2013). The shape model derived from Dawn FC data collected during approach is used, coupled with the spacecraft trajectory and pointing data in NASA's Navigation and Ancillary Information Facility SPICE kernels, to calculate the local illumination and scattering geometry. Some details of the photometric modeling are available in Li et al. (2015).

Due to the limited observing geometry available in our data, especially the lack of data at small phase angles to constrain the opposition effect, some modeled photometric parameters for Occator bright spots, such as the single-scattering albedo (SSA) and geometric albedo, highly depend on the assumptions in the modeling as well as the particular form of photometric models used. The Bond albedo, on the other hand, is much less model-dependent because it is dominated by the light scattered towards moderate range of phase angles (30º-90º) that is well sampled in the data.

Despite relatively large uncertainties in some modeled parameters, photometric models show that the bright spots have significantly different properties from the average Ceres. The SSA of the brightest part of Occator bright spots is between 0.67 and 0.80, comparing to the average SSA of Ceres of 0.094-0.11. Occator bright spots have a Bond albedo of 0.24±0.01 through the Dawn FC clear filter, comparing to 0.034±0.001 for Ceres average. Since SSA and Bond albedo measure the scattered light towards all directions, they are more fundamental physical properties than geometric albedo, which is often poorly constrained due to the lack of data at low phase angle. Hence we compared Occator bright spots with other solar system objects in terms of their Bond albedos. The Bond albedo of Occator bright spots is higher than the average Vesta and most asteroids in the main belt, but is comparable to or slightly lower than the bright material on Vesta (Li et al. 2012) (Fig. 3). Compared to icy moons in the outer solar system, Occator brightest spot is comparable or brighter than Ganymede and Callisto, but darker than Europa and most large Saturnian satellites whose surfaces are dominated by water ice. Therefore, Occator bright spots are unlikely pure water ice, but either admixtures of water ice, salts, and perhaps other bright minerals with darker background material.

The bright spots appear to scatter light more isotropically, with a single-term Henyey-Greenstein parameter of -0.1±0.05, than the average Ceres surface (-0.35±0.05), which is consistent with most asteroids (C- and S-types). The phase function of a surface is affected by, among other factors, grain transparency, shape, defects, impurities, etc. The relatively less backscattering phase function of the bright spots is consistent with higher albedo and stronger multiple scattering, and possibly more transparent particles. The Hapke model roughness parameter for the bright spots is between 40º and 50º, significantly higher than the average Ceres of 20º±3º. Relatively higher roughness suggests more loosely packed regolith particles that cast substantial shadows compared to the average Ceres, consistent with a geologically young regolith that has not been substantially processed (Schröder et al. 2013b).

### 2.3.2 Albedo variability of Occator bright spots

For comparisons of the brightness between Occator bright spots and Haulani crater in different datasets, the effects of different resolution and the vastly different point-spread function (PSF) have to be accounted for. From the radiometrically calibrated images of RC1 and RC3, first, we downsampled them to the pixel scale of HST images (30 km/pixel), then convolve the images with a PSF of HST ACS/HRC at F555W filter generated by the TinyTIM (Krist et al. 2011). We then measured the intensity value of the brightest pixel that contains the Occator crater and Haulani crater from both HST images and the processed Dawn FC images. After correcting for the varying local illumination and viewing angles for both features in all measurements using the best-fit photometric parameters of Ceres surface, we plot the brightness measurements with respect to local solar time for both features and compare their relative brightness (Fig. 2b). The effect of subpixel shift in the downsampling process is negligible.

Ideally, after the correction for local illumination and viewing angle with the best-fit photometric model, the brightness of both features should be independent of local solar time. The residual variations with local solar time as shown in Fig. 2b suggest that the photometric model does not completely remove the dependence on (varying) local scattering geometry over a Ceres day for both features. The slightly different illumination and observing geometries in different datasets result in different trends as they rotate over the disk of Ceres. However, note that the variations after correction in Fig. 2b are at the 4% level, within the performance expectation of photometric correction in general.

Fig. 2b shows that the reflectance contrast between the Occator region and the Haulani region remains unchanged within 2% in HST, and Dawn FC RC1 and RC3 images. The reason that Occator crater appeared to be darker than Haulani region in HST images, as well as in the processed RC1 and RC3 images (Fig. 2b), is due to the large footprint of HST pixels and the wide PSF (80% energy encircled at 5 pixel; Maybhate et al. 2010) of the ACS/HRC camera, together washing out the high contrast of the relatively small bright spots (<4 km) with respect to the background. The bright Haulani crater region, together with the bright rays outside of the crater, are >50 km across. Fig. 2b also shows that no brightness variation was observed in Occator between the RC1 and RC3 observations over the span of eleven weeks. We will continue monitoring the brightness of the bright spots as well as all possible active regions on Ceres throughout the Dawn mission at resolutions down to 35 m/pixel to search for any possible change.

## 3. Discussions

To interpret the putative albedo variability on Ceres we calculated the area in the Occator bright spots and how much albedo change is needed to create a measurable change in global albedo of Ceres. Our albedo variation investigation is limited by the HST pixel scale of 30 km. The Occator bright spots have a combined area of about 95 km$^2$, accounting for 11% of the HST pixel footprint. Therefore our study suggests that any brightness change within the bright spots, or change in their total area, is smaller than 15%.

Given the water production rate of $2\times10^{26}$ molecules s$^{-1}$ for Ceres (Küppers et al. 2014), assuming a Bond albedo of 0.034, and a thermal emissivity of 0.9, pure water ice sublimation model (Cowan & A'Hearn 1979) suggests an active area of 0.2 km$^2$ for a slow rotator case where the surface is at instantaneous thermal equilibrium with local solar insolation at the latitude of Occator (~20º). The low thermal inertia of Ceres of <15 J m$^{-2}$ s$^{-0.5}$ K$^{-1}$ (Chamberlain et al. 2009) justifies the slow rotator assumption. This active area corresponds to 0.2% of the combined surface area of the Occator crater spots, or 0.5% of the central core bright spot. The active area within the bright spots is too small to produce measurable albedo changes at 30 km pixel size.

The impact ejecta observed on main belt asteroid (596) Scheila during the impact in 2010 (Ishiguro et al. 2011) produced an average brightness change of ~5% over 10,000 km$^2$ area. If we use the crater size scaling law (Housen & Schmid 1983) with an exponent parameter of 0.7, based on an escape velocity of 55 m s$^{-1}$ for Scheila (as calculated by Ishiguro et al. 2011) and 0.5 km s$^{-1}$ for Ceres, the same impact on Ceres could produce an ejecta field that is $10^4$ smaller, or ~1 km$^2$. Scaling with albedo and size shows that such an impact is still too small to produce any albedo change on Ceres detectable by our analysis. Based on Bodewits et al. (2011) consideration, the projectile on Scheila had a size of 35-60 m and a relative velocity of ~5.3 km s$^{-1}$; and the impact frequency is once every $10^3$ years. Therefore, if the outgassing on Ceres were triggered by sporadic impact events similar to that considered for Scheila, then such events would not directly produce albedo changes with ejecta near the impact site detectable here.

Perna et al. (2015) suggested that Ceres displays short-term (months) visible spectral variability, and attributed it to "extended resurfacing processes such as cryovolcanism or cometary activity". Our interpretation of the ground-based visible wavelength spectral data shows an increasing spectral slope with increasing phase angle (Fig. 1). Observation geometry has been shown to change slope and band depth in reflectance spectra of small bodies (Sanchez et al. 2012), including Vesta (Reddy et al. 2012, Li et al. 2013). Hence our observations do not support the results reported in that study. The current intermittent and weak water sublimation does not cause albedo and color variations on the surface of Ceres detectable in our data.

Our analysis of the heliocentric dependence of activity on Ceres (Fig. 4) shows no correlation between the two. This is an indication that whatever mechanism responsible for water vapor outgassing on Ceres is at least not completely driven by solar insolation as for comets. The presence of a network of cracks (Reddy et al. 2015b) inside craters with bright spots such as Occator hints at an internal source that is responsible for the transportation of volatiles from their subsurface sources.

We thank the Dawn operations team for the development, cruise, orbital insertion, and operations of the Dawn spacecraft at Ceres. The Framing Camera project is financially supported by the Max Planck Society and the German Space Agency, DLR. Li is supported by a subcontract from the University of California, Los Angeles under the NASA Contract #NNM05AA86 Dawn Discovery Mission. This research made use of Astropy (Astropy Collaboration et al. 2013), and Matplotlib (Hunter 2007)

**Figure Captions**

**Figure 1.** a) The visible spectra of Ceres taken at increasing phase angles (marked at the left of each spectrum) from bottom to top (vertically shifted for clarity). From bottom to top: Geo: Modeled geometric albedo spectrum from ground-based photometric observations (Reddy et al. 2015); P01, P02, P03: Perna et al. (2015) spectra on December 18 and 19, 2012; V87: Vilas et al. (1998) spectrum in 1987; P04, P05, P06: Perna et al. (2015) spectra on January 18, 19, 20, 2013; SOS: $S^2OS^3$ (Lazzaro et al. 2006); RC1: Dawn RC1; SII: SMASS II (Bus & Binzel 2003); V92: Vilas et al. (1998) spectrum in 1992; and RC2: Dawn RC2. b) The linear spectral slopes of Ceres' spectra between 0.54 and 0.85 μm. The colors of symbols correspond to the colors in panel a. The large error bar for the red symbol at 0.8º phase angle, corresponding to P02 spectrum, is due to the much higher noise in the spectrum compared to others. The black dashed line is a linear fit to the spectral slope with respect to phase angle, not including the slope measured from Dawn RC2 at 44º phase angle. The spectral slope of Ceres increases nearly linearly from 0º to 25º phase angle.

**Figure 2.** a) Images of Haulani crater (left column) and Occator crater (right column) from HST taken in December 2003 and January 2004 (top row), Dawn RC1 images taken on February 4, 2015 (second row), Dawn RC3 images taken on May 5, 2015 (third row), and the downsampled and convolved images of Dawn RC3 images (bottom row). The to craters in all images are marked by the dashed-line circles, and their names are noted in the RC3 images. The HST images shown in the top row are drizzled (Fruchter & Sosey 2009) to 0".015 per pixel scale from the original 0".025. The pixel scale at Ceres shown in the top row is 18 km for HST images, for RC1 images 13.7 km, for RC3 images 1.32 km, and for processed RC3 images 30 km. SC and SS are the longitude and latitude of, respectively, sub-spacecraft/HST and sub-solar points; and ph is the phase angle. b) The photometrically corrected reflectance of Haulani region (red) and Occator region (blue) in three datasets. The reflectance are measured from original HST images, and Dawn RC1 and RC3 images that are convolved with HST/ACS/HRC PSF and downsampled to HST image resolution. The I/F are corrected to a common, arbitrary geometry of 3.15º incidence angle, 3º emission angle, and 6.15º phase angle, close to the phase angle of HST images. The horizontal lines are the averages of the I/F for both features to show the contrast between the two regions in various cases. The contrast between Haulani region and Occator region at 30 km scale is 5-7%, and remain unchanged within 3% between HST, RC1, and RC3.

**Figure 3.** Comparisons of Bond albedos between Ceres (red square), Occator bright spots (blue square), and other solar system objects (green circles). The Bond albedos of icy satellites are from a list complied by Verbiscer et al. (2012); and those of comets (9P, 19P, 81P, and 103P), asteroids (asteroid numbers 4, 243, 253, 433, 951, 2867, and 26143), the Moon, and the Martian satellites are from a list compiled by Li et al. (2013).

**Figure 4.** The positions of Ceres in its orbit during the previous searches of water vapor. The solid ellipse is Ceres' orbit with the Sun at the center, and "Peri" marks the perihelion of Ceres. The filled symbols mark the detection of water vapor (circles, by Herschel) or OH (triangles, by IUE), and the open symbols are non-detection. The directions of the triangles mark the direction of off-limb OH search, where upper triangles are off northern limb, and down triangles are off southern limb. The S and N are summer solstice for southern hemisphere and northern hemisphere, respectively, based on the pole measurement from Dawn data. The blue shaded part of Ceres' orbit is during Dawn's planned stay from April 1$^{st}$, 2015 to June 30$^{th}$, 2016. No obvious trend of water outgassing and orbital phase (perihelion/aphelion, seasons) is clearly identified. The outgassing appears to be intermittent and occurs at arbitrary orbital phases.

**Table 1.** Observation conditions extracted from the FC color data.

|  | Phase Angle (deg) | Average resolution (km/pixel) | S/C distance (km) | Heliocentric distance (AU) |
|---|---|---|---|---|
| RC1 | 17.2 – 17.6 | 7.68 | 82,151 | 2.8529 |
| RC2 | 42.7 – 45.3 | 4.24 | 45,359 | 2.8578 |
| RC3 (Equatorial) | 7.7 – 11.0 | 1.28 | 13,603 | 2.9051 |

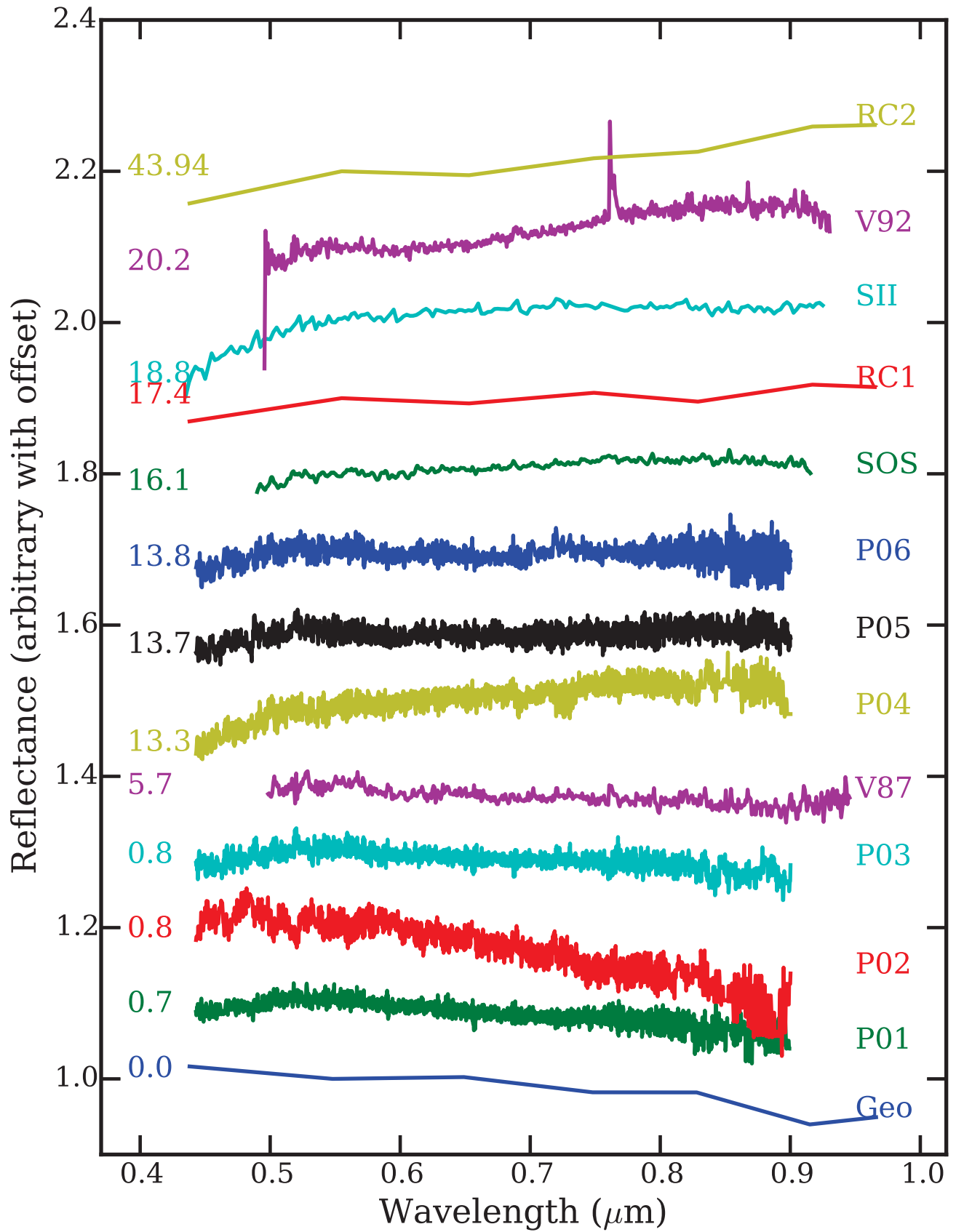

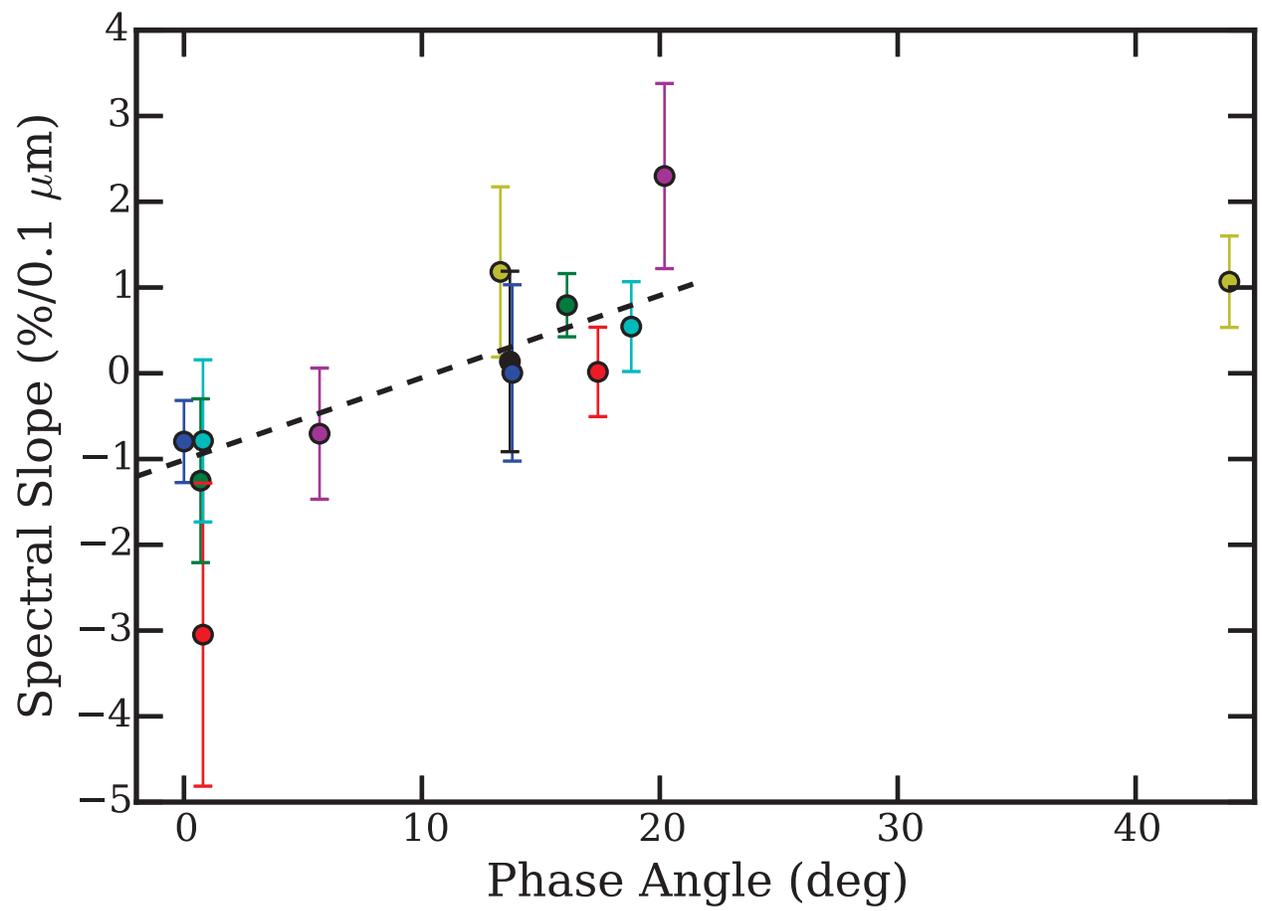

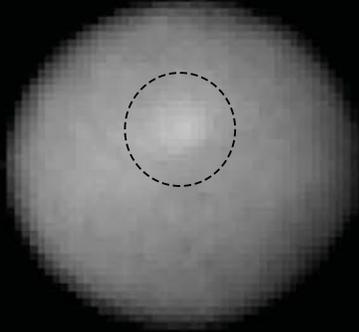 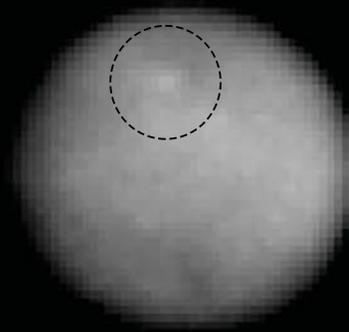
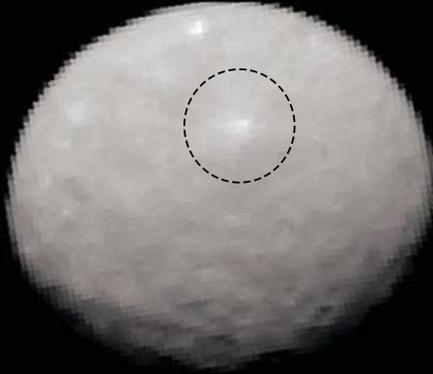 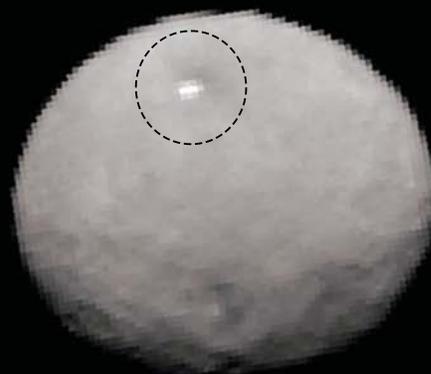
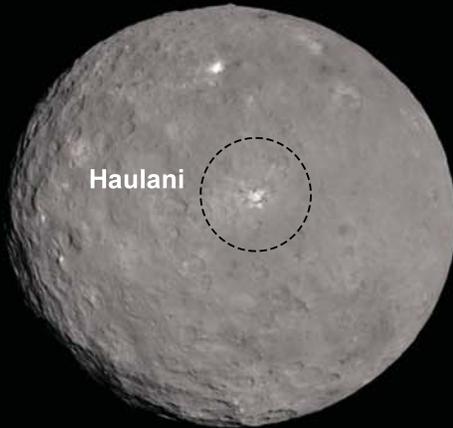 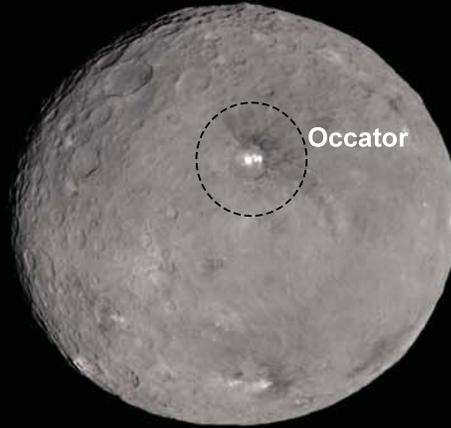
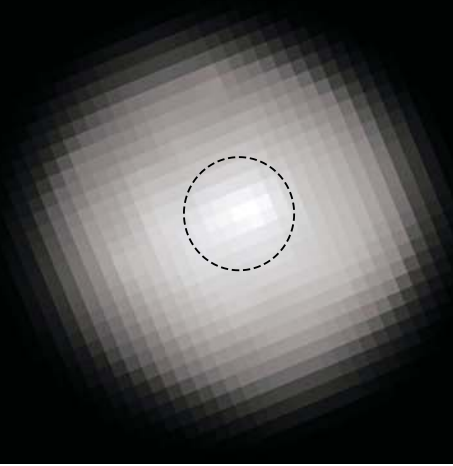 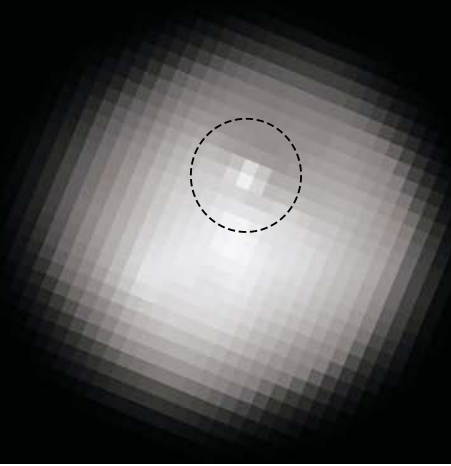

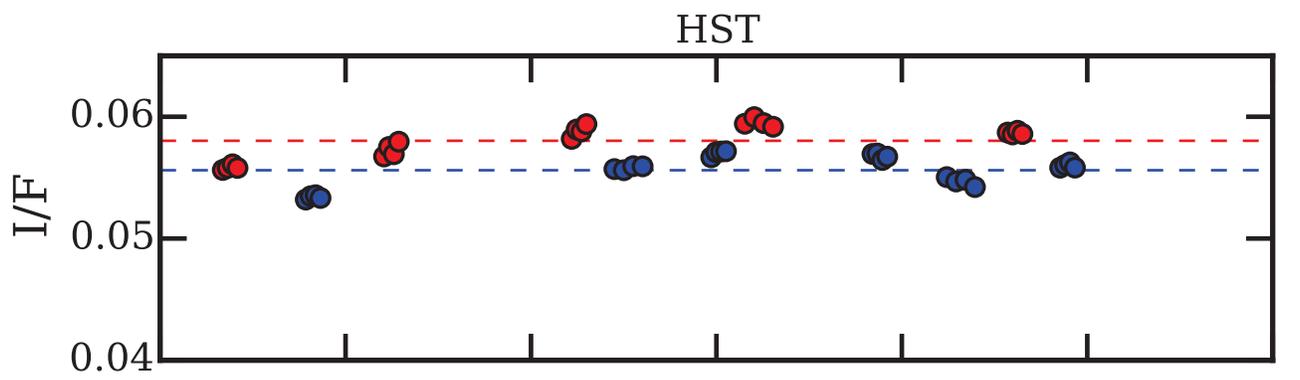
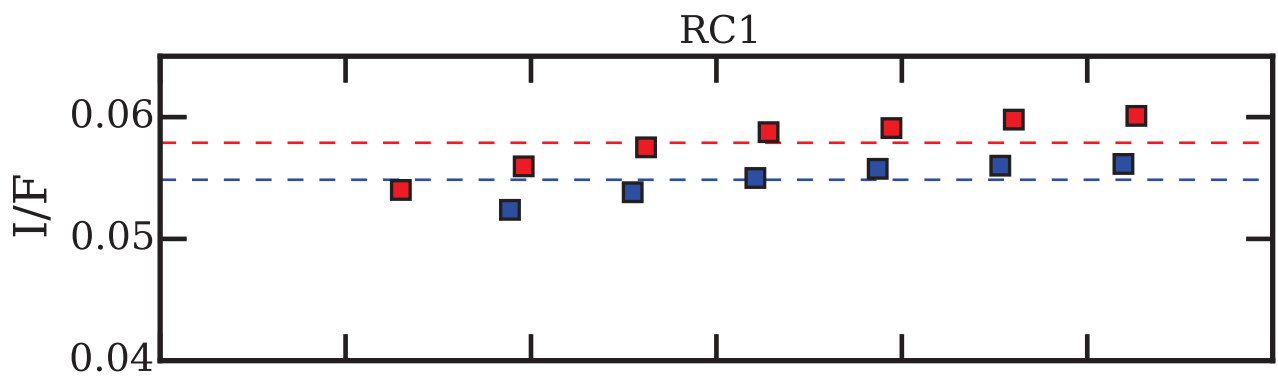
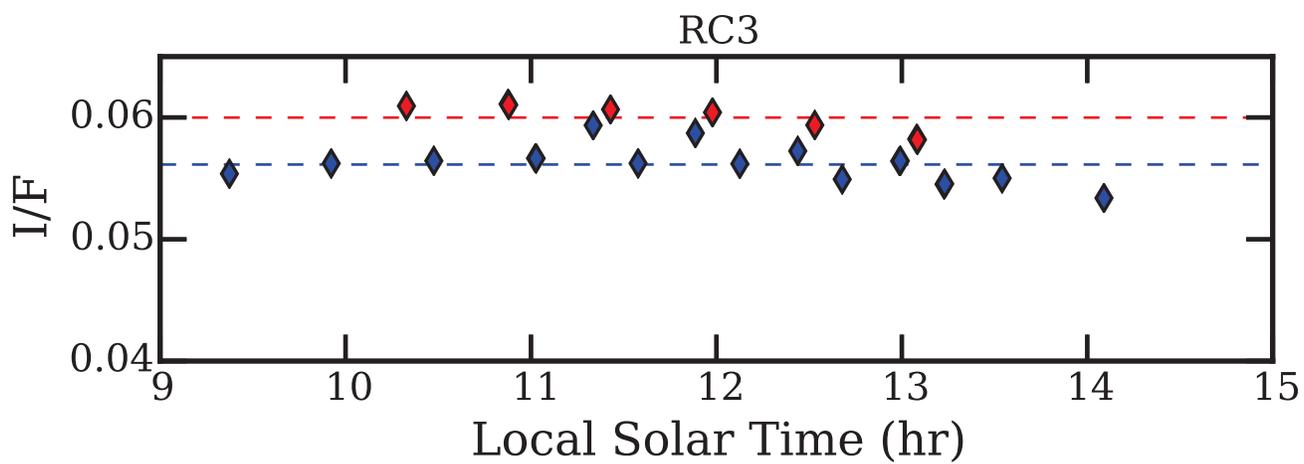

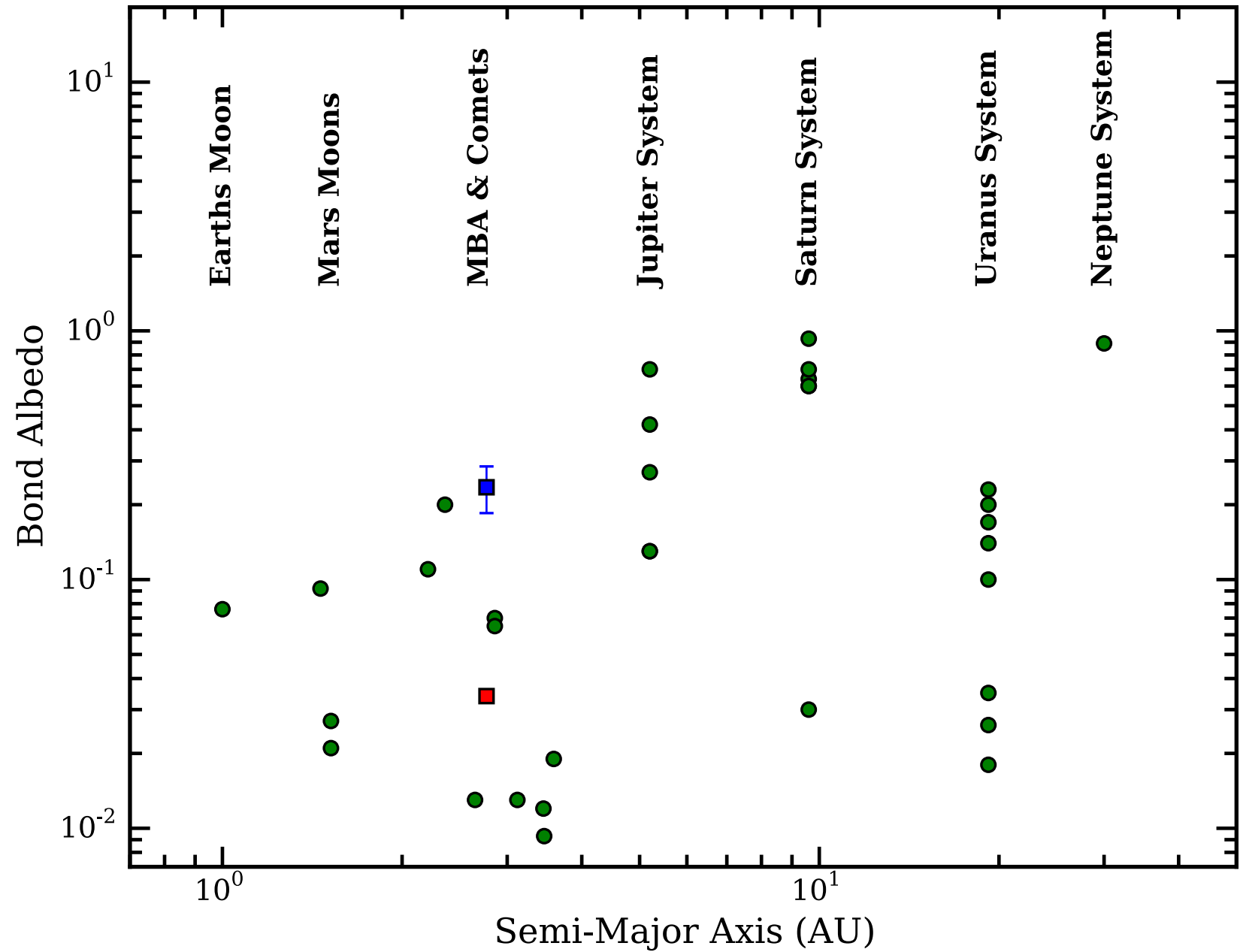

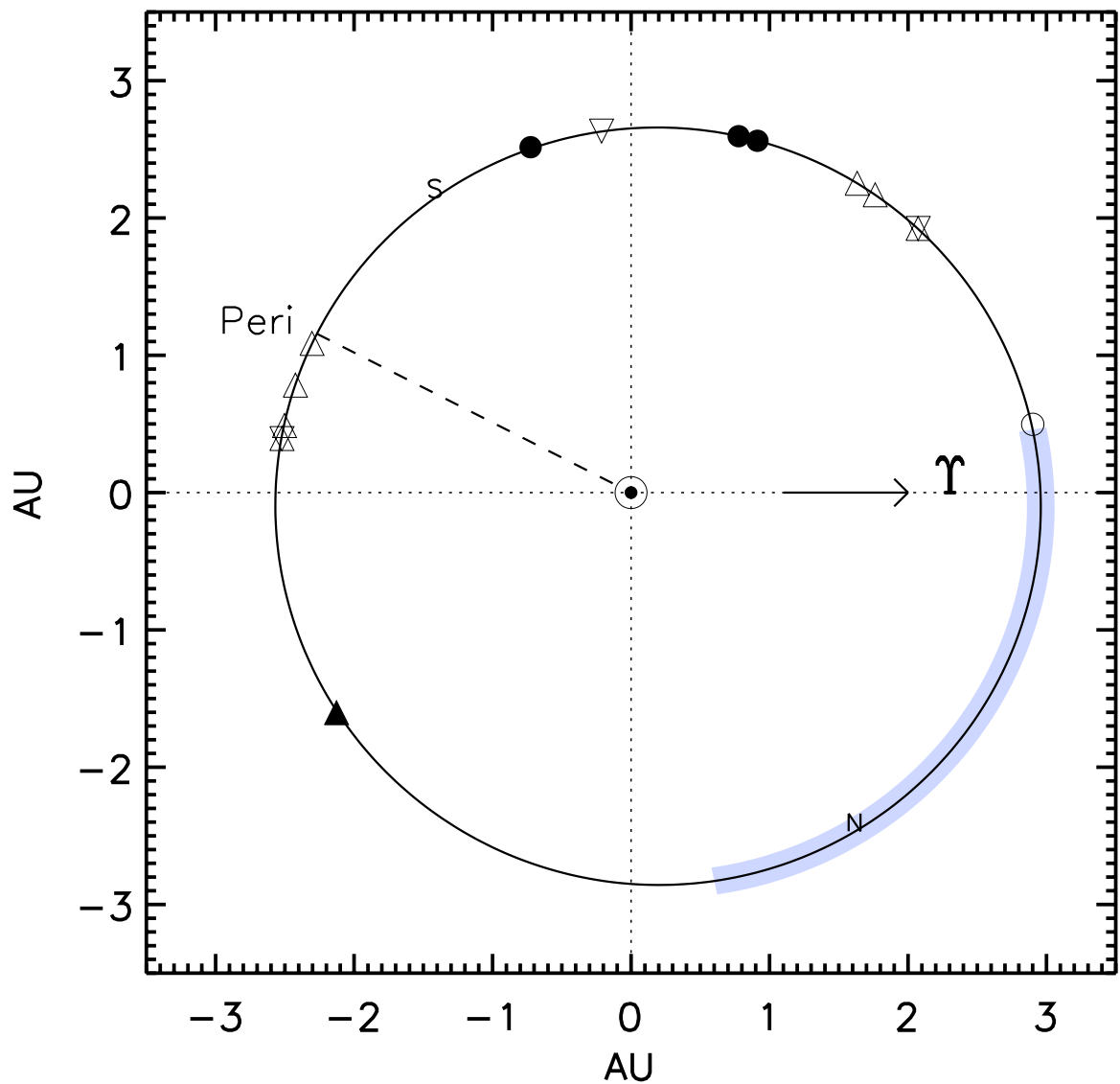